\documentclass[10pt,aps,prl,twocolumn,superscriptaddress]{revtex4}

\usepackage[utf8]{inputenc}
\usepackage{hyperref}
\usepackage{graphicx}
\usepackage{amsmath,amssymb}
\usepackage{color}
\usepackage{subfigure}
\usepackage{flushend}

\graphicspath{{Images/}}

\begin{document}

\title{
Holographic laser Doppler imaging of pulsatile blood flow
}

\author{Jeffrey Bencteux}
\author{Pierre Pagnoux}
\author{Thomas Kostas}

\affiliation{
Centre National de la Recherche Scientifique (CNRS) UMR 7587, Institut Langevin. Fondation Pierre-Gilles de Gennes. Institut National de la Sant\'e et de la Recherche M\'edicale (INSERM) U 979, UPMC, Universit\'e Paris 7. Ecole Sup\'erieure de Physique et de Chimie Industrielles ESPCI ParisTech - 1 rue Jussieu. 75005 Paris. France
}

\author{Sam Bayat}
\affiliation{
Universit\'e de Picardie Jules Verne. INSERM U 1105. Paediatric Lung Function Laboratory, Amiens University Hospital, Amiens, France.
}

\author{Michael Atlan}

\affiliation{
Centre National de la Recherche Scientifique (CNRS) UMR 7587, Institut Langevin. Fondation Pierre-Gilles de Gennes. Institut National de la Sant\'e et de la Recherche M\'edicale (INSERM) U 979, UPMC, Universit\'e Paris 7. Ecole Sup\'erieure de Physique et de Chimie Industrielles ESPCI ParisTech - 1 rue Jussieu. 75005 Paris. France
}

\date{\today}
%\date{}

\begin{abstract}
We report on wide-field imaging of pulsatile motion induced by blood flow using heterodyne holographic interferometry on the thumb of a healthy volunteer, in real-time. Optical Doppler images were measured with green laser light by a frequency-shifted Mach-Zehnder interferometer in off-axis configuration. The recorded optical signal was linked to local instantaneous out-of-plane motion of the skin at velocities of a few hundreds of microns per second, and compared to blood pulse monitored by plethysmoraphy during an occlusion-reperfusion experiment.
\end{abstract}

\maketitle

\section{Introduction}\label{sect:intro}

One of the most commonly displayed clinical waveforms is the finger plethysmogram. It allows the non-invasive measurement of pulse wave amplitude, which reflects the changes in finger blood flow. It is usually recorded by a pressure sensor~\cite{GanongBarrett2005} or a pulse oximeter which illuminates the skin and measures changes in optical absorption~\cite{HertzmanSpealman1937, Allen2007, AlianShelley2014}. Optical plethysmography is widely available, its feasibility has even been demonstrated using mobile phones~\cite{JonathanLeahy2010, ScullyLeeMeyer2012}. However, it does not provide a spatially-resolved image. This limitation can be overcome by signal processing of regular video recordings in ambient light for time-dependent spatial and color fluctuations enhancement~\cite{TanakoOhta2007, ScaliseBernacchia2012, WuRubinsteinShih2012, BalakrishnanDurandGuttag2013}. Other imaging approaches are designed to harness optical phase fluctuations of coherent laser light from the motion of body walls, either by single-point scanning detection~\cite{EssexByrne1991}, direct image detection of temporal~\cite{Draijer2009, Leutenegger2011, KamshilinTeplov2013} and spatial speckle contrast~\cite{BoasDunn2010, HumeauHeurtierMaheDurandAbraham2013}, or holographic interferometry~\cite{LeclercqKarrayIsnard2013, DobrevFurlong2014}.\\ 

In this letter, we report on narrow band laser Doppler imaging of superficial pulsatile motion of the thumb by heterodyne holographic interferometry with a camera refreshed at 60 Hz. A transient blood flow interruption experiment in a healthy volunteer is performed and compared to blood pulse monitoring. A major advantage of this system is that no physical contact with the studied tissue surface area is required. This could allow assessment of non-intact skin or mucous membranes or even the intraoperative study of tissues within the surgical field. Moreover, the lack of surface contact reduces the risk of infection transmission which is a true concern with reusable contact-requiring biomedical equipment~\cite{Wilkins1993}, also avoiding the extra cost of disposable contact probes.\\  

\begin{figure}[b]
\centering
\includegraphics[width = 7.0 cm]{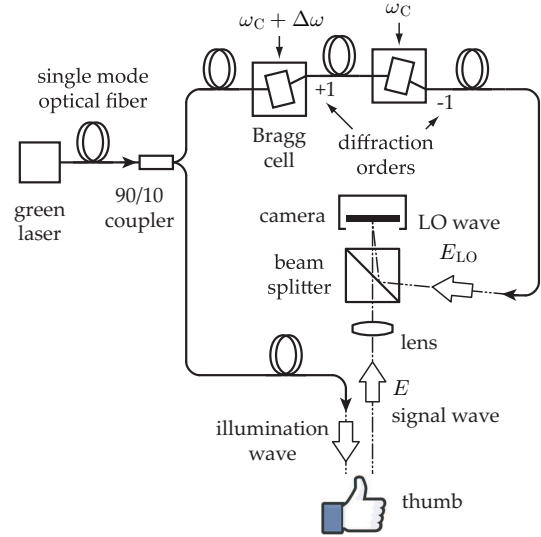}
\caption{Sketch of the fibered Mach-Zehnder heterodyne holographic interferometer. The main laser beam is split into two channels. In the object channel, the optical field $E$ is backscattered by the skin. In the reference channel, the optical field $E_{\rm LO}$ is frequency-shifted by two Bragg cells by $\Delta\omega$. A standard camera records interferograms of the diffracted fields $E$ and $E_{\rm LO}$.}
\label{fig_Setup}
\end{figure} 

\section{Experimental setup}\label{sect:Setup}

The experimental laser Doppler imaging scheme designed for this study is sketched in Fig.~\ref{fig_Setup}. The apparatus consists of a fibered Mach-Zehnder optical interferometer for off-axis~\cite{LeithUpatnieks1962} and frequency-shifting~\cite{Aleksoff1969, Macovski1969} holographic interferometry. The main optical radiation field is provided by a 150 mW, single-mode, fibered diode-pumped solid-state green laser (Cobolt Samba-TFB-150) at wavelength $\lambda = 532$ nm, and optical frequency $\omega_{\rm L} / (2 \pi) = 5.6 \times 10^{14} \, \rm Hz$. The thumb is illuminated with about $\sim$ 10 mW of continuous optical power, over $\sim 40 \, {\rm mm} \times 40 \, {\rm mm}$. In the reference channel, an optical local oscillator (LO) is formed by optical frequency shifting. Two acousto-optic modulators (Bragg cells, AA Opto Electronic), driven with phase-locked signals at $\omega_{\rm C}$ and $\omega_{\rm C} + \Delta \omega$, are used to shift the optical frequency of the laser beam from $\omega_{\rm L}$ to $\omega_{\rm L} + \Delta\omega$. The carrier frequency $\omega_{\rm C}/(2\pi)$ is set to 200 MHz, at the peak response of the fibered acousto-optic modulators. The backscattered optical field $E$ is mixed with the LO field $E_{\rm LO}$ with a non-polarizing beam splitter cube. A Ximea MQ042MG-CM camera records interference patterns at a frame rate of $\tau_{\rm S}^{-1} = \omega_{\rm S} / (2 \pi) = 60 \, \rm Hz$. Each raw interferogram of $2048 \times 2048$ pixels, digitally acquired at time $t$ is noted $I(t) = \left| E(t) + E_{\rm LO}(t) \right| ^2$. A small angular tilt $\sim 1 ^\circ$ between $E$ and $E_{\rm LO}$ ensures off-axis recording. The sensor is set $\sim 50 \, \rm cm$ away from the object plane. A convergent lens of 50 mm focal length is present in the object channel, in order to widen the lateral field of view of the holographic detection to $\sim 50 \, {\rm mm}$. The backscattered laser optical field $E(t) = {\cal E}  \exp \left[ i \omega_{\rm L} t + i \phi(t) \right]$ undergoes a phase variation $\phi(t)$, as a consequence of out-of-plane motion of the illuminated tissue. It is mixed with the LO field from the reference channel $E_{\rm LO}(t) = {\cal E}_{\rm LO} \exp \left[ i \omega_{\rm L}t + i \Delta \omega  t \right]$, that is tuned to a close-by intermediate frequency $\omega_{\rm L} + \Delta \omega$. The quantities ${\cal E}$ and ${\cal E}_{\rm LO}$ are complex constants and $i$ is the imaginary unit. The magnitude of a given optical Doppler component is retrieved by frequency down conversion within the camera's temporal bandwidth, ensured by non-linear detection of the field $E$ by the array of square-law sensors of the camera, that respond linearly with the optical irradiance $I(t)  =  \left| E(t) + E_{\rm LO}(t) \right|^2$. The squared magnitude of the total field received, $E + E_{\rm LO}$, has cross-terms oscillating at the difference frequency of the fields $E$ and $E_{\rm LO}$
\begin{equation}\label{eq_I}
I(t) = \left| {\cal E} \right|^2 + \left| {\cal E}_{\rm LO} \right|^2 + H(t) + H^*(t)
\end{equation}
where $H(t) = {\cal E} {\cal E}_{\rm LO}^*\exp \left( i \phi - i \Delta \omega t \right)$ is the heterodyne interferometric contribution, and $^*$ denotes the complex conjugate. This equation describes the temporal fluctuation of the recorded irradiance at a given pixel. 

\section{Signal processing}\label{sect:SignalProcessing}

Holographic image rendering, or spatial demodulation, is then performed onto each recorded interferogram $I$ with a discrete Fresnel transform~\cite{Goodman1967}.  In off-axis recording configuration, The Fresnel transform separates spatially the four interferometric terms of the right member of Eq.~\ref{eq_I}~\cite{LeithUpatnieks1962, Cuche2000}. After spatial demodulation of each interferogram, the heterodyne signal $H(t)$ appears in the off-axis region of the hologram, processed as follows : First, the squared amplitude of the difference of two consecutive off-axis holograms is formed in order to cancel very low frequency noise contributions 
$S^2 = \left|H(t) - H(t-\tau_{\rm S})\right|^2$. Then, to cancel laser intensity fluctuations, a normalization factor $\left<S_0^2\right> = \left< \left|H_0(t) - H_0(t-\tau_{\rm S})\right|^2\right>$ is formed. The contribution $H_0$ is measured in a region of interest of the reconstructed hologram where the terms of Eq.~\ref{eq_I} are not present~\cite{AtlanGross2006}, and the brackets $\left< \, \right>$ account for spatial averaging. In high heterodyne gain regime, i.e. when $\left|{\cal E}\right|^2 \ll \left|{\cal E}_{\rm LO}\right|^2$, the quantity $\left< S_0^2 \right>$ is dominated by shot-noise of the LO~\cite{GrossAtlan2007, LesaffreVerrier2013}, which scales up linearly with $\left|{\cal E}_{\rm LO}\right|^2$. The averaged signal $\left< S^2 \right>$ scales up linearly with $\left|{\cal E}\right|^2 \left|{\cal E}_{\rm LO}\right|^2$. Hence the processed Doppler signal $\left< S^2 \right>/\left< S_0^2 \right>$ is a heterodyne measurement of the optical power (or irradiance) $\left|{\cal E}\right|^2$, which does not depend on the LO power~\cite{MagnainCastelBoucneau2014}.

\section{Narrowband apparatus response}\label{sect:ApparatusResponse}

\begin{figure}[]
\centering
\includegraphics[width = 8.0 cm]{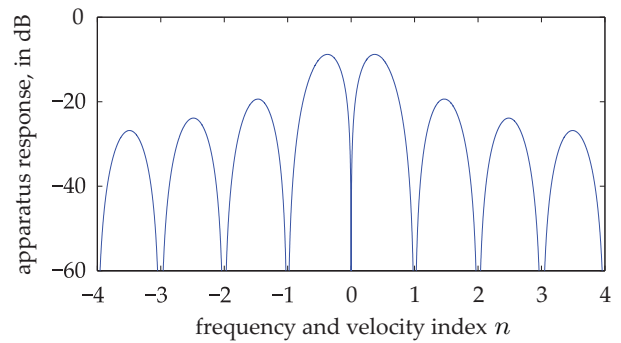}
\caption{Apparatus response (Eq.~\ref{eq_fPSF}), calculated for $\tau_{\rm S} = \tau_{\rm E}$, versus apparent (beating) frequency index $n = \omega/\omega_{\rm S}$. The ticks of the horizontal axis correspond to the zeros of the response at integer values of the index $n$ of the beating frequency $n\omega_{\rm S}$ and the probed velocity $v_n = (\Delta\omega + n \omega_{\rm S})/(2k)$.}
\label{fig_ApparatusResponse}
\end{figure}

Two-phase demodulation of interferograms recorded with a sampling frequency approximately equal to the reciprocal of the exposure time ($\tau_{\rm S}^{-1} \simeq \tau_{\rm E}^{-1}$) results in a narrowband detection with an apparatus response plotted in Fig.~\ref{fig_ApparatusResponse}. This response filters-off optical contributions which are not shifted in frequency, i.e. statically-scattered light. For the backscattered field's irradiance $\left|{\cal E}\right|^2$, this response is~\cite{AtlanDesbiolles2010, VerrierAlexandreGross2014}
\begin{equation}\label{eq_fPSF}
B(\omega)  = \frac{1}{\omega^2 \tau_{\rm E}^2}\sin^2 \left( \frac{\omega\tau _{\rm S}}{2} \right)\sin^2 \left( \frac{\omega \tau_{\rm E}}{2} \right) 
\end{equation}
where $\omega$ is the apparent frequency recorded by the camera, as a result of heterodyne beating.

Let's assume that at the short time scale of the sampling process $t < \tau_{\rm S} = 2\pi/\omega_{\rm S}$, the local instant velocity is approximately constant. $v(t)\approx v(t -\tau_{\rm S}) \approx v$. In other words, its magnitude $v$ is identified to the time-averaged value of the instant velocity during $\tau_{\rm S}$, so we can simply write the local optical phase variation due to the transverse displacement of the skin as $\phi(t) = 2 k v t = \omega t$, when the illumination and the scattered wave vectors are perpendicular to the skin's surface. The optical wave number is $k = 2\pi/\lambda \simeq 1.2 \times 10^{7} \, \rm rad.s^{-1}$, and $\omega = 2kv$. Now let's consider temporal variations of the instant velocity of norm $v(t)$ at long time scales $t > \tau_{\rm S}$. The light reflected by the skin undergoes a local time-dependent Doppler shift
\begin{equation} \label{eq_DopplerShift}
\omega(t) = 2 k v(t)
\end{equation}
To assess a given instant velocity $v_0$, the LO is detuned by $\Delta\omega = 2 k v_0$. According to Eq.~\ref{eq_DopplerShift} and Eq.~\ref{eq_fPSF}, it results in the detection of Doppler-shifted light with a velocity-dependent efficiency $B\left(2k(v-v_0)\right)$. The range of probed velocities $v$ within the two first lobes of the apparatus response $B$ is
\begin{equation}\label{eq_v_range}
[v_{-1}, v_{+1}] = \left[v_0 - \frac{ \omega_{\rm S}}{2k}, v_0 + \frac{\omega_{\rm S}}{2k} \right]
\end{equation}
\begin{figure}[]
\centering
\includegraphics[width = 8.0 cm]{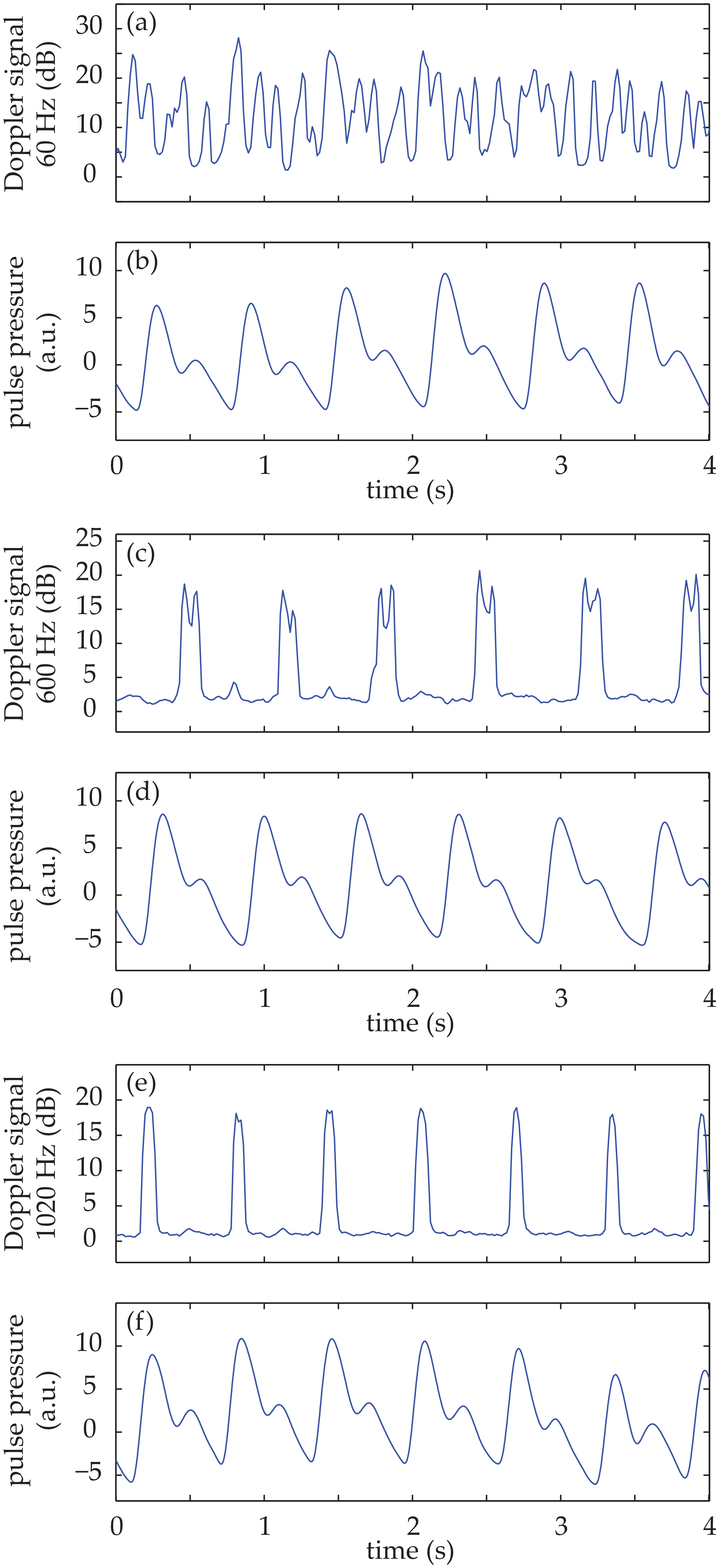}
\caption{Doppler signal (a,c,e), averaged in the central part of the image of the thumb (Fig.~\ref{fig_ThumbDoppler1020Hz}), and concurrent blood volume signal (b,d,f), versus time. Detuning frequencies : $\Delta \omega / (2 \pi) = 60 \, \rm Hz$ (a), $\Delta \omega / (2 \pi) = 600 \, \rm Hz$ (c), $\Delta \omega / (2 \pi) = 1020 \, \rm Hz$ (e). The synchronization accuracy of holographic and standard blood volume measurements is of the order of $\pm 0.5\, \rm s$. a.u.: arbitrary units.}
\label{fig_DopplerVsBloodVolume4s}
\end{figure} 
\begin{figure}[]
\centering
\includegraphics[width = 8.0 cm]{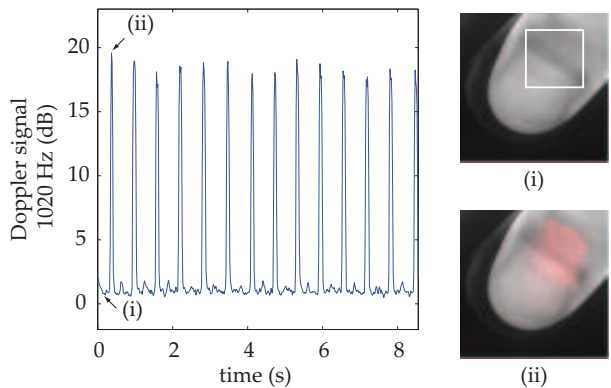}
\caption{Optical Doppler signal, acquired for a detuning frequency $\Delta \omega / (2 \pi) = 1020 \, \rm Hz$, averaged in the region of interest depicted in (i), versus time. Composite images of the corresponding signal at low (i) and high (ii) signal level. Red color indicates signal presence. The whole sequence is reported in \href{http://youtu.be/kGPWyUIM1OA}{in this video}.}
\label{fig_ThumbDoppler1020Hz}
\end{figure} 
\begin{figure}[]
\centering
\includegraphics[width = 8.0 cm]{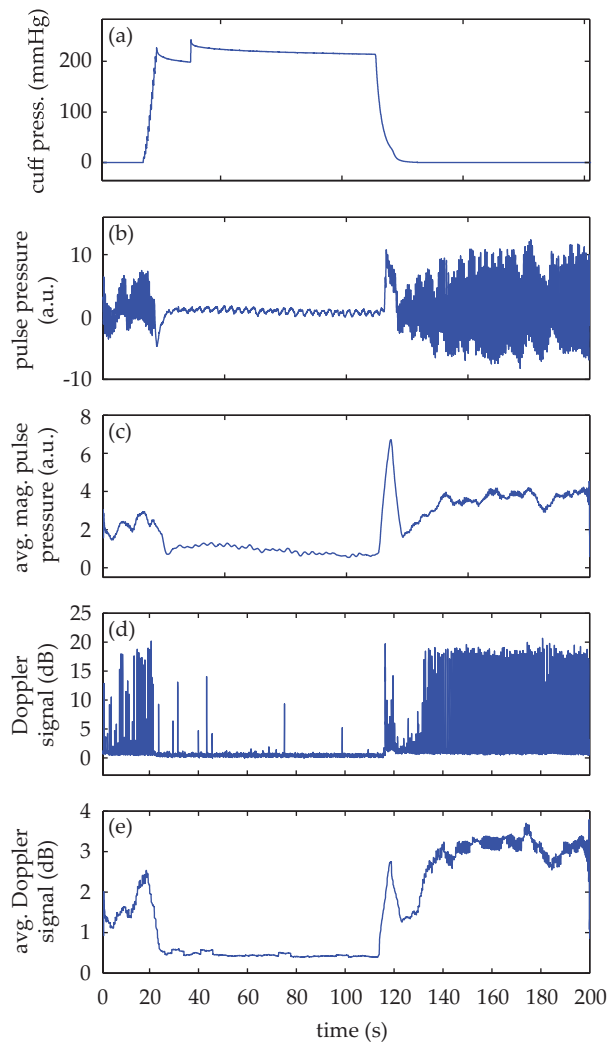}
\caption{Time traces of indicators monitored during a hypoperfusion experiment. Cuff pressure (a), plethysmogram [raw data (b), 5 s time-average of the magnitude (c)], holographic laser Doppler signal recorded at 1020 Hz and averaged spatially in the nail area [raw data (d), 5 s time-average (e)]. A time-lapse sequence of the composite Doppler image of the thumb is reported \href{http://youtu.be/DddF2QAjhcQ}{in this video}.}
\label{fig_PulseVolumeVsDoppler1020Hz}
\end{figure} 

\section{Experiment}\label{sect:Experiment}

An experiment on the fingers of healthy volunteers undergoing arterial occlusion provoked with a pressure cuff was conducted. The pressure cuff was placed around the right arm. The cuff bladder pressure was measured using a transducer (MLT0670, ADInstruments, Oxford, UK). The signal was amplified, digitized at 200 Hz using an analog/digital converter (Powerlab; ADI Instruments), and recorded on a computer. The arterial pulse wave was monitored using a piezo-electric pulse transducer (TN1012/ST, ADInstruments) placed on the index of the imaged hand. The signal was recorded at 200 Hz. The thumb was illuminated by the green laser beam. A comfortable support for the forearm was used to avoid stray motion. A preliminary experiment was performed to monitor the Doppler response at three different detuning frequencies $\Delta \omega / (2 \pi)$ : $60 \, \rm Hz$, $600 \, \rm Hz$, and $1020 \, \rm Hz$; choosing a detuning frequency which is an integer number of times the sampling frequency is not mandatory, but is has the advantage of canceling efficiently statically-backscattered light, according to the apparatus response (Fig.~\ref{fig_ApparatusResponse}).  These frequencies correspond to probed velocities of $16\, \rm \mu m.s^{-1}$, $160\, \rm \mu m.s^{-1}$, and $271\, \rm \mu m.s^{-1}$, respectively. Fig.~\ref{fig_DopplerVsBloodVolume4s} shows plots over time of pulse waveforms (plethysmograms) and holographic Doppler signal $10 \log_{10}\left(\left< S^2 \right>/\left< S_0^2 \right>\right)$ (in dB) averaged in a region of interest, placed on the central part of the image of the thumb. At very low frequency detunings ($\lesssim 100 \, \rm Hz$), motion artifacts prevented the measurement of pulsatile motion. When the frequency was set to $\Delta\omega/(2\pi) = 1020 \, \rm Hz$, which corresponded to a measurement range of instant velocities $[v_{-1}, v_{+1}] = [255\,\mu {\rm m .s} ^{-1}, 287\,\mu {\rm m .s} ^{-1}]$, centered at $v_0 = \Delta\omega / (2 k) = 271 \,\mu {\rm m .s} ^{-1}$ (Eq.~\ref{eq_v_range}), the Doppler signal appeared to be less prone to motion artifacts. The Doppler signal at 1020 Hz recorded over $\sim$ 8 s, and composite images of the regionalized Doppler response (represented in red) are displayed in Fig.~\ref{fig_ThumbDoppler1020Hz} and \href{http://youtu.be/kGPWyUIM1OA}{in this video}. This value of the frequency shift was chosen for the blood flow interruption experiment, reported in Fig.~\ref{fig_PulseVolumeVsDoppler1020Hz}. The cuff pressure and four indicators were monitored : pulse wave, 5 s-averaged pulse wave magnitude, Doppler signal, and 5 s-averaged Doppler signal. Blood flow to the forearm was occluded by increasing the cuff pressure to $\sim$ 200 mmHg during a $\sim$ 90 s time interval. During occlusion, both recorded signals dropped. Short spikes due to motion artifacts were observed on the Doppler signal; they were more pronounced at lower Doppler frequency. On releasing the cuff pressure, a transient increase of both the pulse volume and the Doppler signal is observed as expected, due to post-occlusive hyperemia. A time-lapse sequence of the composite Doppler image of the thumb during the occlusion-reperfusion experiment is reported \href{http://youtu.be/DddF2QAjhcQ}{in this video}. A limitation of holographic Doppler imaging of pulsatile blood flow with visible laser light is its sensitivity to low velocities, which lets motion artifacts prevent the measurement of pulsatile motion, and requires stabilization of the monitored tissue. This issue might be alleviated with infrared light.

\section{Conclusion}\label{sect:Conclusion}

In conclusion, we demonstrated wide-field imaging and monitoring of pulsatile motion of the thumb of a healthy volunteer. Narrowband optical Doppler video imaging in real-time was performed with a off-axis and frequency-shifting holographic interferometer. Numerical image rendering was performed by discrete Fresnel transformation and two-phase temporal demodulation at a 60 Hz framerate. The measured contrast was linked to the instant velocity of out-of-plane motion, of the order of a few hundreds of microns per second. Robust non-contact motion monitoring was achieved. This was illustrated by an occlusion-reperfusion experiment, in which the proposed detection scheme was compared to plethysmography.

\section*{Acknowledgments}\label{sect:Conclusion}

We gratefully acknowledge the contributions of Anna Kieblesz, Renaud Boistel, Aurore Bonnin, and Claude Boccara, and financial support from Agence Nationale de la Recherche (ANR-09-JCJC-0113, ANR-11-EMMA-046), Fondation Pierre-Gilles de Gennes (FPGG014), r\'egion Ile-de-France (C’Nano, AIMA), the Investments for the Future program (LabEx WIFI: ANR-10-LABX-24, ANR-10-IDEX-0001-02 PSL*), and European Research Council (ERC Synergy HELMHOLTZ).

\end{document}